\documentclass[a4paper, amsfonts, amssymb, amsmath, preprint, showkeys, nofootinbib, twoside]{revtex4-1}
\usepackage[english]{babel}
\usepackage{subfigure}
\usepackage[utf8]{inputenc}
\usepackage[colorinlistoftodos, color=green!40, prependcaption]{todonotes}
\usepackage{amsthm}
\usepackage{mathtools}
\usepackage{physics}
\usepackage{xcolor}
\usepackage{graphicx}
\usepackage[left=23mm,right=13mm,top=35mm,columnsep=15pt]{geometry} 
\usepackage{adjustbox}
\usepackage{placeins}
\usepackage[T1]{fontenc}
\usepackage{lipsum}
\usepackage{csquotes}
\usepackage[pdftex, pdftitle={Article}, pdfauthor={Author}]{hyperref} 

\def\be{\begin{equation}}
\def\ee{\end{equation}}
\def\ber{\begin{eqnarray}}
\def\eer{\end{eqnarray}}
\def\bwt{\begin{widetext}}
\def\ewt{\end{widetext}}
\def\tr{\textcolor{red}}

\begin{document}
\title{A novel mechanism for probing the Planck scale with wavepackets following general distributions
}

\author{Saurya Das}
    \email{saurya.das@uleth.ca}
    \affiliation{Theoretical Physics Group and Quantum Alberta, Department of Physics and Astronomy, University of Lethbridge, 4401 University Drive, Lethbridge, Alberta T1K 3M4, Canada}
\author{Sujoy K. Modak}    
    \email{smodak@ucol.mx}
    \affiliation{Facultad de Ciencias - CUICBAS, Universidad de Colima, Colima, C.P. 28045, M\'exico}

\date{\today} 

\begin{abstract}

The Generalized Uncertainty Principle (GUP) is predicted by most theories of quantum
gravity and in turn introduces a minimum measurable length in nature. 
%
It was also shown recently that GUP predicts potentially measurable corrections to the `doubling time' of freely moving Gaussian atomic and molecular wavepackets with a favorable combination of three parameters, {\it e.g.} mass, initial width and mean velocity of a travelling wavepacket. 
However, it is well known that such wavepackets can come with various shapes which correspond to variety of distributions. In this article, we generalize our earlier work for an {\it arbitrary distribution} and thereby accommodate any shape of the wavepacket. 
Mathematically, we build this formalism by exploiting a duality between quantum and statistical mechanics, by which (quantum mechanical) expectation values of the momentum operator can be expressed in terms of the derivatives of the characteristic functions of the dual statistical description.
Equipped with this result, we go one step further and numerically study a few physical distributions. We find that large organic (TPPF152) wavepacket following the generalized normal distribution with parameter $\kappa=0.5$ offers one of the best-case scenarios, effectively scanning the whole GUP parameter space with current technologies. 
Although we do not say that the minimal length has to be near or at the Planck value, we mange improving our previous studies to scan the minimal length signatures down to hundred times the Planck value.




\end{abstract}


\maketitle

\tableofcontents

\section{Introduction}

The idea of a fundamental Minimal Length Scale (MLS) in Nature has surfaced over several decades \cite{planck,tomil,hberg,yang,deser,maed1,maed2,garay}. In addition, a modification of the Heisenberg Uncertainty Principle (HUP) by the so called Generalized Uncertainty Principle (GUP) is often advocated from the point of view of MLS as well as from a diverse set of quantum gravity studies \cite{gup1,gup2,gup3,kmm,kpp,golam,gup4,scard,smolin,Amel,dv1,adv,ghosh,bdm,doug,hossen,ac}. 
 GUP has two clear advantages, namely, its simplicity makes it easy to use and perform calculations with a given quantum mechanical system and the second is a possibility of experimental verification of the results obtained by considering GUP effects. However, in practice, calculations in standard quantum mechanical systems that have been historically tested for the confirmation of the standard quantum theory have almost negligible GUP effects beyond any measurement. Therefore, recent theoretical works with motivations for testing GUP have advocated for new avenues where these tiny effects can be measurable. Like many phenomenological models, GUP also comes with an undetermined parameter (often denoted by a dimensionless parameter $\alpha_0$). In \cite{Villalpando:2019usm} it was shown that $\alpha_0$ can be connected with the MLS and that would imply a hard bound $1\le \log_{10}(\alpha_0) \le 16$, between the GUT scale and the Planck scale\footnote{Valid for GUP with both linear and quadratic terms.}. Determination of the value of $\alpha_0$ is important since this will specify a value for a fundamental MLS in Nature, in a phenomenological manner. Proposals for determination of the GUP parameter is an important task and there exist several estimations/bounds of the same by various works which are divided into two classes in \cite{values} -- non-gravitational and gravitational contexts \footnote{Although some of the bounds were noted for the quadratic GUP parameter $\beta_0$, it is straightforward to transfer those bounds to $\alpha_0$, simply because $\beta_0\sim \alpha_0^2$.}. Within the non-gravitational context, precision measurement of Landau levels and Lamb shift provide an upper bound $\log_{10}(\alpha_0) \le 25$ and  $\log_{10}(\alpha_0) \le 18$  \cite{dv1, adv} which turns out to be outside the hard limit set by \cite{Villalpando:2019usm}. Further, considering heavy mesons as harmonic oscillators, another bound can be obtained for the GUP parameter which is in accordance with the upper bound of  \cite{Villalpando:2019usm}. On the other hand, considering macroscopic harmonic oscillators, the upper bound is found to be much lower $\log_{10}(\alpha_0) \le 6$ \cite{bawaj}. In addition, considering a gravitational framework, there is a proposal to fix the value of the GUP parameter, as found to be $\alpha_0=\sqrt{82\pi/5}$ \cite{bhole}. 
 
 While various approaches provide bounds on the GUP parameter we are interested to consider a novel path for probing the GUP parameter all the way down to the Planck scale. Our previous papers laid out a foundation of this path showing  that quantum QG/GUP signatures  via the GUP may be measurable in the laboratory by studying the evolution of quantum wavepackets \cite{Villalpando:2018xsh}-\cite{Das:2021yqn}.  In particular, it was demonstrated that the time taken for the width of a Gaussian wavepacket to be double its starting value, or in other words the `doubling time', is affected by the GUP/Planck scale parameters. To provide a basic outline of a possible experiment  we recall following two well known experimental studies: first, time resolved dynamics of the nuclear wave-packets of He-He* dimer and Ne$_2$ dimer ions were studied in \cite{6,7} and,  on the other hand, recently, large molecular wave-packets were used in interference experiments, such as in \cite{dbs3}  where in total of 430 atoms simultaneously reach the left and right slits, separated by a distance 2 to 3 times larger than the size of the wave-packets. These later wave-packets pass through the slits and expand during their journey before showing interference patterns on the screen. Mindful of the above two types of experiments performed already, here we propose a thematic experiment integrating the above two (i.e., a time-resolved experiments with multi-atomic macromolecular wavepacket). 

It is worth mentioning that the GUP  implies a modification of the 
standard energy-momentum dispersion which may give rise to the so-called composition law problem or the soccer-ball problem. It is a situation where energies and momenta of the constituents of a composite body do not simply add up, even for non-interacting constituents. The magnitude of GUP effects depends on whether it is applied to the constituents or the centre-of-mass of the system \cite{hossen,ac}. This remains an open issue. 
Although there exist approaches which try to address this issue 
\cite{rgup,2scales}, 
here we adopt the viewpoint of \cite{pik1,pik2,bdm}, that one first defines the quantum system under consideration (in our case the expanding wavepacket) and applies GUP to the system and estimate quantum gravity corrections therein.
It should be clear that here we apply GUP to the well-defined and experimentally verified quantum wavepacket of the entire system of atoms. The latter may or may not be deducible from the individual wavepackets due to their complicated mutual interactions. Furthermore, even if we work with the individual wavepackets as suggested in \cite{hossen} we do not face soccer-ball problem to a measurable amount. This is because for typical momenta of atoms considered here
($\approx 10^{-22}$kg-m/s ) and the relatively small number of atoms in these system ($\approx 600-700$), the maximum estimated (relative) error due to the potential soccer-ball problem is tiny ($\approx 10^{-19}$ for each unit of GUP deviation)
and therefore can be safely ignored.  
Eventually, it is up to experiments to determine the correctness of the approach.

We would like to mention that there is a complementary proposal of introducing the minimal length scale without modifying the commutator bracket (i.e., GUP), but  rather by modifying the momentum operator \cite{doug23, doug24} itself. 
Such an approach may be useful in the context of certain astrophysical observations \cite{astrogup}.
 
%
In our earlier works \cite{Villalpando:2018xsh}-\cite{Das:2021yqn} a numerical algorithm was used, whose aim was to quantify certain physical variables, whose correct combinations can make these effects much larger and potentially measurable with current technology, no matter how small is the GUP parameter. These variables are the mass, initial width and the mean velocity of a quantum mechanical wavepacket. It turns out that greater the mass, initial width and mean velocity, greater is the `doubling time'. Our estimate suggested that with the best available large quantum wavepacket of Gaussian shape, and corresponding to TPPF152 molecule (in terms allowed combinations of above three variables) may be able to scan all the way down to $\log_{10}\alpha_0 \sim 5$ which then scan every bound set in non-gravitational contexts. However, we are still away from the gravitationally suggested value which sets $\log_{10}\alpha_0 \sim 1$.

As mentioned above, our previous estimates of probing the GUP parameter by measuring the doubling time with the state of the art accuracy was calculated only for Gaussian wavepackets, which are difficult to prepare and maintain during the evolution process. Thus, simply from the practical point of view, it is necessary to allow as many kinds of shapes as possible beyond the Gaussian. 
%
%
It therefore has two clear advantages: (a) we would be able to test the robustness of earlier results, i.e., their dependency on the shape or distribution of the wavepacket which might encourage, or to the contrary, rule out the necessity of measuring the `doubling time', and (b) to see if further improvements of the estimated corrections can be made, simply by manipulating various shapes/distributions of wavepackets, so that we can scan the GUP parameter space beyond the non-gravitational bounds and to even probe the value proposed by heuristic gravitational arguments. 

In this paper, we show that both the above can be addressed, with encouraging results, if we extend our work to include a diverse set of distributions, beyond just the Gaussian that was chosen in previous works \cite{Villalpando:2018xsh}-\cite{Das:2021yqn}.
Here, we provide a precise recipe to compute the Planck scale correction for the wavepacket expansion rate, especially practical for the likes of a composite system such as a large molecule, and
following practically any distribution, as long as it has a finite variance. 
The latter condition is automatically satisfied of course, on physical grounds.
This is achieved by expressing all quantum mechanical (and potentially measurable) results in terms of characteristic functions of a chosen distribution. This is possible because the results involve various moments of the momentum of the wave-packets, which in standard quantum mechanics, is evaluated using the expectation values of the momentum operator and its powers. However, if a statistical description is available for the quantum system of interest, such as for the multi-particle quantum mechanical wave-packets, the same moments have a dual description in terms of the Characteristic Functions (CF) of the associated statistical distribution describing the wave-packet dynamics. 
Here we exploit this `quantum-statistical duality', and build an appropriate  bridge to bypass various quantum mechanical complexities (which might also be a priori unavailable) and use a relatively well-known and straightforward tool of statistical mechanics. Expressing the $n^{th}$ moment of the momentum operator in terms of the $n^{th}$ derivative of the CF gives us this tool.\footnote{Another nearly equivalent route could be constructed using the Kolmogorov formula since CFs of finite variance has a dual description using the Kolmogorov formula. However, if one follows that procedure oneself will end up dealing with $n^{th}$ order integration using appropriate kernels which renders the procedure much more lengthy and complicated \cite{Reichl}.} We show that this is indeed a powerful machinery, by applying our results to the specific examples of 
(a) the Gaussian, (b) the Poisson and (c) the generalized normal distributions. As a cross-check, the results of the previous works \cite{Villalpando:2018xsh}-\cite{Das:2021yqn} are recovered using the CF procedure. In addition, from the latter more general distributions, an appropriate limit is shown to exist which provides the results of the Gaussian distribution as reported in \cite{Villalpando:2018xsh}-\cite{Das:2021yqn}.  
This reinforces the usefulness of our method and its potential applicability to a large number of distributions and provides us in total of {\it four parameters -- mass, initial width, mean velocity and shape of the wavepacket} to induce GUP effects inside the reach of latest atomic clocks. 

This paper is organized a follows. In the next Section \ref{sec2} we give a brief overview of GUP and relativistic corrections to the dynamics of free wavepackets. In 
Section \ref{sec3} we build a general framework for interpreting and expressing various quantum mechanical results with statistical mechanical tools. We express the broadening rate in terms of the CFs in section \ref{sec4}. In the next section \ref{sec5} we show the equivalence of the quantum mechanical and statistical approaches for the Gaussian distribution. The modified expansion rates for two more general distributions (beyond Gaussian) are calculated in section \ref{sec6}.  Finally, in section \ref{sec7} we compute numerically the quantum gravity corrections to the doubling time using the expressions for the CFs and compare the results with state of the art accuracies of time measurements. Finally, we conclude in Section \ref{sec8}.

\section{Broadening of free wavepackets}\label{sec2}

We start with the Hamiltonian for a free particle of mass $m$ in $(1+1)$-dimensions, including the leading order relativistic correction term
\begin{eqnarray}
H
&=& \frac{p^2}{2m} - \frac{p^4}{8m^3c^2} \label{ham1} 
\end{eqnarray}
Now, as per GUP, the fundamental commutator between position and momentum is modified to 
\cite{adv}
\begin{eqnarray}
[x,p] = i\hbar\, [1-2\alpha p +4 \alpha^2p^2]~.
\end{eqnarray}
The above defines a minimum measurable length and a maximum measurable momentum, in terms of the GUP parameter $\alpha$. The minimum measurable length is given by the minimum value of the position uncertainty \cite{Villalpando:2019usm}
\begin{equation}
 (\Delta x)_{\text{min}} 
 = \frac{3\,\alpha_0}{2}\,\ell_{Pl}.
 \label{alpha}
\end{equation}
Correspondingly, there is also a maximum uncertainty in momentum, given by \cite{Villalpando:2019usm}
\begin{equation}
 (\Delta p)_{\text{max}} 
 = 
 \frac{M_{Pl}\,c}{2 \alpha_0}.
\label{alpha11}
\end{equation}
Note that, the maximum uncertainty in momentum implies that one cannot have an infinite momentum in the GUP picture since for an infinite momentum, the corresponding uncertainty may also be infinite, which would clearly contradict \eqref{alpha11}.
Here we have defined  
$\alpha= \alpha_0/M_{Pl} c$, $\alpha_0$ being dimensionless. 
$M_{Pl}$ is the Planck mass, 
$M_{Pl} c$ the Planck momentum, 
$M_{Pl} c^2 \approx 10^{16}$ TeV the Planck energy and 
$\ell_{Pl} \approx 10^{-35}$ m is the Planck length. 
We do not assume any specific value of $\alpha_0$, but hope that experiments will shed light on the allowed values of $\alpha_0$. 
Since no evidence of a MLS has been found in high energy experiments such as the LHC, one is automatically led to an upper bound on $\alpha_0$. 
Together with a lower bound on it corresponding to the Planck scale, one arrives at the following allowed range:
$1 \leq \alpha_0 \leq 10^{16}$. Note that, since $M_{Pl} c$ is of the order of unity in SI units, 
the same numerical bounds apply to the dimensionful GUP parameter $\alpha$ as well.

 
%

Next, we define an auxiliary momentum variable 
$p_0$, which is `canonical' in the sense that $[x,p_0]=i\hbar$. Therefore as an operator, one can write $p_0=-i\hbar\,d/dx$. 
This is related to the physical (i.e. measurable) momentum $p$ via the relation $p= p_0(1-\alpha p_0 + 2 \alpha^2 p_0^2)$. Substituting in Eq.(\ref{ham1}), one obtains the following effective Hamiltonian for a relativistic system, incorporating GUP
{
\begin{eqnarray}
%
%
H= H_{\text{NR}} + H_{\text{rel}} + H_{\text{LGUP}} + H_{\text{QGUP}} + H_{\text{LGUP}}^{\text{rel}}
\label{hfull}
\end{eqnarray}
where,
(i) $H_{\text{NR}} = \frac{p_0^2}{2m}$,
(ii) $~H_{\text{rel}} = -\frac{p_0^4}{8m^3c^2}$,
(iii) $~H_{\text{LGUP}} = - \frac{\alpha}{m} p_0^3,$
(iv) $~H_{\text{QGUP}} = \frac{5\alpha^2}{2m} p_0^4$,
and (v) $~H_{\text{LGUP}}^{\text{rel}} = \frac{\alpha}{2m^3c^2}p_0^5$. 
In the above, 
(i) is the standard non-relativistic Hamiltonian, 
(ii) the leading order relativistic correction, 
(iii) the linear GUP correction (proportional to $\alpha$), 
(iv) the quadratic GUP correction 
(proportional to $\alpha^2$)
and (v) the hybrid or mixed term,
which includes both the relativistic and linear GUP correction.  

Next, we study the evolution of free wavepackets with the above Hamiltonian. 
It is well-known that a free wave-packets broadens, i.e. their widths increase with time due to the Heisenberg's uncertainty principle. One can, for example, use the Ehrenfest theorem to estimate this broadening. 
Here our interest is to consider the modified broadening rate   of the free wave-packet with the full Hamiltonian with GUP corrections \eqref{hfull} \footnote{Any potential gravitational decoherence effect is ignored as we are interested in purely GUP modifications. Such decoherence effects, even if they exist, are expected to be minuscule.}. As is well-known, the Ehrenfest's theorem gives the time derivative of the expectation values of the position 
($x$) and its canonically conjugate momentum ($p_0$) operators as follows:
$\frac{d}{dt} \langle x \rangle = \frac{1}{i \hbar} \langle [ x, H ] \rangle = \Big\langle \frac{\partial H}{\partial p_0} \Big\rangle$ 
and
 $\frac{d}{dt} \langle p_0 \rangle = \frac{1}{i \hbar} \langle [p_0, H] \rangle = - \Big\langle \frac{\partial H}{\partial x} \Big\rangle$. These can be extended to the expectation of any operator of course, and in particular to $p_0^n$, which appear in (\ref{hfull}) for various integer values of $n$. 
For the above, one obtains
 $\frac{d}{dt} \langle p_0^n \rangle = \frac{1}{i \hbar} \langle [ p_0^n, H ] \rangle =0$, implying that 
 $\langle p_0^n \rangle = \text{constant in time}$.
 
%


Next, to estimate the DTD, we first write the first and second time-derivatives of the 
square of the width (or variance) of the quantum mechanical wave-packet, which is 
defined as $\xi = \Delta x^2 = \langle x^2\rangle -\langle x \rangle^2$:
\ber
\dot{\xi} &=& \frac{d\xi}{dt} = \frac{d}{dt} \langle x^2 \rangle - 2 \langle x \rangle \frac{d  \langle x\rangle}{dt} \label{dxi}\\
\ddot{\xi} &=& \frac{d^2\xi}{dt^2} = \frac{d^2}{dt^2} \langle x^2\rangle - 2 \left(\frac{d  \langle x \rangle}{dt}\right)^2 - 2x \frac{d^2 \langle x \rangle}{dt^2}~. \label{ddxi}
\eer
The above can be simplified using the Ehrenfest theorem and the Hamiltonian given in (\ref{hfull}). 
%

Following the detailed analysis of \cite{Das:2021yqn}, one can 
re-write the above as follows

%
%
\be
 \ddot{\xi}_{\text{full}} = \frac{2}{m^2} \Delta p_0^2 
 + C_{\text{rel}} + C_{\text{LGUP}} + C_{\text{QGUP}} + C_{\text{LGUP}}^{\text{rel}}, \label{masteq}
\ee
where
\ber
C_{\text{rel}} &=& -\frac{2}{m^4c^2} (\langle p_0^4\rangle - \langle p_0 \rangle \langle p_0^{3} \rangle)+ \frac{\Delta {p_0^{(3)}}^2}{2m^6c^4} \label{cr1} \\
C_{\text{LGUP}} &=& -\frac{12\alpha}{m^2} (\langle p_0^3\rangle - \langle p_0 \rangle \langle p_0^{2} \rangle) + \frac{18\alpha^2}{m^2} {\Delta {p_0^{(2)}}^2} \label{cr2} \\
C_{\text{QGUP}} &=& \frac{40\alpha^2}{m^2} (\langle p_0^4\rangle - \langle p_0 \rangle \langle p_0^{3} \rangle) +\frac{200\alpha^4}{m^2} {\Delta {p_0^{(3)}}^2} \label{cr3} \\
C_{\text{LGUP}}^{\text{rel}} &=& \frac{2\alpha}{m^3c^2} (\langle p_0^5\rangle - \langle p_0 \rangle \langle p_0^{4} \rangle) +\frac{25\alpha^2}{2m^6c^4} {\Delta {p_0^{(4)}}^2} \label{cr4}
\eer
all are defined at the initial time $t=0$.

Using the definition of $\xi$, the master equation \eqref{masteq} has the following solution, which gives the rate of broadening of the free wavepacket under the combined influence of the relativistic and GUP corrections
\begin{widetext}
\be
\Delta x (t) = \sqrt{\xi_{\text{in}} + \dot{\xi}_{\text{in}} t + \frac{(\Delta p_0^2)_{\text{in}}}{m^2} t^2  + \frac{1}{2} \left( C_{\text{rel}} + C_{\text{LGUP}} + C_{\text{QGUP}} + C_{\text{LGUP}}^{\text{rel}}\right)t^2},
\label{freegup1}
\ee
\end{widetext}
where, the subscript ``$\text{in}$'' corresponds to the initial value of the various quantities, such as the initial width ($\sqrt{\xi_{\text{in}}}$), the initial rate of expansion  $\dot{\xi}_{\text{in}}$ and the initial variance of the canonical momentum $(\Delta p_0^2)_{\text{in}}$, and new corrections due to the relativistic and GUP effects appearing in \eqref{masteq}.
In the rest of the paper, we will consider two important additions to our previous works \cite{Villalpando:2018xsh}-\cite{Das:2021yqn}, namely: \\
(i) we will consider non-minimal wavepackets, and \\
(ii) we will study distributions other than Gaussian,
\\
and show that the bulk of our results continue to hold and further widens the window of opportunity for measuring Planck scale effects in the laboratory, in terms of a wider range of physical systems and parameters.

\section{An invitation to statistical mechanics}\label{sec3}

Most textbook and research studies focus
overwhelmingly on Gaussian wave-packets. 
While this may simplify the theory side, it greatly
restricts the freedom to test various wave-packets on the experimental side.
In this section, we will precisely go beyond the 
Gaussian wave-packet approximation.
We will consider a wave-packet following an
{\it arbitrary} statistical distribution,
which more accurately describes 
an experimental set up meant to 
study wave-packets and their broadening rates, by means of time resolved experiments. 
This requires an understanding of the role of statistical mechanics in our study, and in particular \eqref{freegup1}, which we present below .

\subsection{The term {$\dot{\xi}_{\text{in}}(t)$}}

Before we discuss the details of 
the wave-packet expansion following 
various distributions functions, 
we give a simple analysis for understanding the coefficient of the linear order term in time ($t$) appearing in \eqref{freegup1}. 
For any value of $t$, we can calculate this term using the Ehrenfest's relations, which for any operator $A$, tells
\be\label{eheq}
\frac{d}{dt}\langle A \rangle (t) = \langle [A, H] \rangle + \frac{\partial A}{\partial t},
\ee
where $H$ is the Hamiltonian under consideration. For a free particle wave-packet this is just $p^2/2m$ and using $\xi = \langle q^2 \rangle - \langle q \rangle^2$ we get
\be\label{dxi}
\dot\xi (t) = \frac{1}{m} \left(\langle xp + px \rangle - 2\langle x \rangle \langle p \rangle \right)
\ee
where we have used the classical approximation $\langle p \rangle = m \frac{d}{dt} \langle x \rangle$. Using the commutation relation $[x,p]=i\hbar$, we re-express the above as \footnote{For the HUP, $p$ is the standard/physical momentum, whereas, for the GUP one may consider $p=p_0$, the canonical momentum which is different from the physical momentum in the GUP picture.}
\be\label{dxii}
\dot\xi (t) = \frac{1}{m}\left[i\hbar + 2 (\langle px \rangle - \langle p \rangle \langle x \rangle) \right].
\ee

Now we can we make a connection of the above quantum mechanical equation with statistical mechanics by recalling the distribution of two jointly distributed stochastic variables, $Y$ and $Z$. 
The covariance between $Y$ and $Z$ is given as
\be\label{cov}
\text{Cov}(YZ) = \langle yz \rangle - \langle y \rangle \langle z \rangle
\ee
where $y$ and $z$ are the stochastic outcomes of the respective variables. Setting $Y=\hat{p}$ and $Z=\hat{x}$ then introduces a stochastic interpretation of \eqref{dxii}, which can now be 
re-written as
\be\label{dxicov}
\dot\xi (t) = \frac{1}{m}\left[i\hbar + 2~ \text{Cov}(\hat{p}\hat{x})~ \right].
\ee
In fact, we can also introduce the often used
correlation function between $\hat{p}$ and $\hat{x}$ as
\be
\text{Cor}(\hat{p}\hat{x}) = \frac{\text{Cov}(\hat{p}\hat{x})}{\sigma_p\sigma_x},
\ee
where $\sigma$s are the standard deviation in the respective spaces. 
In terms of the correlation function,
\eqref{dxicov} becomes
\be\label{dxicor}
\dot\xi (t) = \frac{1}{m}\left[i\hbar + 2\sigma_p\sigma_x~ \text{Cor}(\hat{p}\hat{x})~ \right].
\ee
We see therefore that the dynamics of free wave-packets can be written in terms of 
stochastic distributions.  

This discussion gives us a physical interpretation of the term $\dot{\xi}_{\text{in}}(t)$ -- it is a measure of correlation between the momentum and position of the wave-packet at the initial time $t_0$. For a Gaussian wave-packet one can show that $2\sigma_p\sigma_x~ \text{Cor}(\hat{p}\hat{x}) = -i\hbar$, which makes $\dot{\xi}_{\text{in}}(t) = 0$ as is expected for such a wavepacket which is the only one with a minimal width saturating the uncertainty principle{\footnote{One can easily check this using the wavefunction \eqref{gpsi}.}}.
Naturally, for any other distribution this term need not vanish and is dependent on the strength of correlation between the mean position and momentum of the wave-packet. 
Since different distributions of wave-packets lead to  different correlations, the initial value $\dot{\xi}_{\text{in}}(t)$ depends in principle on the experimental template. 
The value however, can be calculated by measuring the overall contraction or expansion of the initial wave-packet at the initial time $t_0$.  
For the initially expanding wave-packet, one should have $\dot{\xi}_{\text{in}}(t)>0$, and in case of a  contracting wave-packet $\dot{\xi}_{\text{in}}(t)<0$. Since the correlation function satisfy the condition  $-1 \le \text{Cor}(\hat{p}\hat{x}) \le +1$ (plus implies correlation and minus anti-correlation), we can obtain an upper bound on the magnitude of $|\dot{\xi}_{\text{in}}(t)|$ which also depends on the product of the standard deviations in position and momentum spaces. This would allow one to infer possible values of the velocity term  $|\dot{\xi}_{\text{in}}(t)|$. In addition, the same term can in principle be measured at any given instant of time, and be  considered as an experimental input. We shall see, however, for the numerical study to be conducted later on in this paper, this term does not affect the outcome to a measurable amount for the cases to be considered in this study. Nevertheless, even to neglect the effect of this term we need to have an appropriate insight on this term and the above discussion does provide us one. 

\subsection{Characteristic functions}

In this subsection we will argue that characteristic functions  play a key role in generalizing our theoretical results to arbitrary distributions and also takes us one step closer 
towards direct comparison with experiments. 

Results of wave-packet broadening reviewed in the last section in \eqref{cr1} - \eqref{freegup1} involve moments of various orders, both in the position and in the momentum space coordinates. We shall show here that the characteristic function (CF), defined in the position space, corresponding to the probability distribution of the momentum operator (in the momentum space), can be used to achieve following two improvements - (i) we can eliminate moments of higher orders (greater or equal to two) in momentum by functions defined in the coordinate space, and (ii) with the use of the CF, we shall have a general expression of the wave-packet broadening that can be applied to the various stochastic distributions in agreement with the spectrum of the momentum operator.

We believe this takes us one step closer to  experiments because as we know, quantum 
measurements are generally carried out in position 
space, and therefore, while comparing any experimental data of wave-packet broadening with our theoretical results, it may be problematic if the results are not completely in position space. 

Now, let us recall the definition of the CF in a stochastic distribution. The CF, $f_Y(x)$, corresponding to the stochastic variable, $Y$, is defined as \cite{Reichl}
\ber
f_Y(x) = \int_{-\infty}^{\infty} dx \, e^{ixp/\hbar} P_{Y}(p) = \sum_{n=0}^{\infty} \frac{(ix)^n \langle p^n \rangle}{\hbar^n\,n!},
\label{cf}
\eer
where $P_Y (p)$ is the probability distribution of the stochastic variable $Y$ in the momentum space. In our case, we set $Y=\hat{p}$, i.e., the momentum operator, for which the set $\{p\}$ consists of all possible stochastic outcome of the eigenvalues of the momentum operator. 
In equation \eqref{cf}, the CF corresponding to the momentum operator in the position space is expressed in terms of the probability distribution of the momentum operator in momentum space. The series expansion is justified only when the higher order moments $\langle p^n \rangle$ are small so that the series is convergent. The CF is a continuous function of $x$ and has following properties --- $f_{\hat{p}} (0) =1$, $|f_{\hat{p}} (x)| \le 1$ and  $f_{\hat{p}}^{*}(x) = f_{\hat{p}}(-x)$. The inverse transformation of \eqref{cf} also exists, given by
\be
P_{\hat{p}} (x)= \frac{1}{2\pi} \int_{-\infty}^{\infty} dp\, e^{-ipx} f_{\hat{p}} (p).
\ee
Using the series expansion in \eqref{cf} we can also obtain the moments by derivating the CF
\ber
\langle p^n \rangle &=& \hbar^n\,\lim_{x\rightarrow0} (-i)^n \frac{d^n f_{\hat{p}}(x)}{dx^n},\\
&=& (-i)^n\, \hbar^n\,
\lim_{x\rightarrow0}  f_{{\hat{p}},n}(x)
\label{mom}
\eer
With these definitions we are in a position to express the GUP broadening rate \eqref{freegup1} in terms of the derivatives of the CF in position space{\footnote{One can instead use the relationship between the CF with Kolmogorov kernel and express the results solely based on the kernel. Recall that,  Kolmogorov formula is a general formula defining the CF for distributions with finite variance \cite{Reichl}, in the form $f_Y(k) = \exp{\left[i\gamma k + \int_{-\infty}^{+\infty} \left(e^{iku} - 1 -iku \right)\frac{dK(u)}{u^2}\right]}$, where $\gamma$ is a constant and the kernel $K(u)$ is  a non-decreasing function with bounded variation. Note  that, $\lim_{k\rightarrow 0}f_Y'(k) = i\gamma$ and $\lim_{k\rightarrow 0}f_Y''(k) = -\gamma^2 - K(\infty)$. Therefore, the first moment $\langle k \rangle = \gamma$ and the standard deviation $\sigma = \sqrt{K(\infty)}$. This procedure based on the choice of the Kernel describing various distributions although is feasible to follow, but, it renders the procedure much more complicated. Therefore, we prefer to use the CF and not the Kolmogorov kernel. }}


\section{Broadening rates involving the characteristic function}
\label{sec4}
Expressions for the time evolution of the width of the wave-packet, governed by the equations \eqref{cr1} - \eqref{freegup1}, involve up to the eighth  order moment $(n=8)$ of momentum $\langle p_0^n \rangle$. It is straightforward to replace all these moments  in terms of the derivatives of the CF by using \eqref{mom}.

First, the standard quantum mechanical contribution appearing both in \eqref{freegup1} can be rewritten as
\ber
\Delta p_0^2 &=& \hbar^2\,\lim_{x\rightarrow 0} \left(f_{{\hat{p}_{0}},1}^2(x) - f_{{\hat{p}_{0}},2}(x) \right) \label{n1}
\eer
which includes up to the second order derivative of the CF. Notice that the initial uncertainty in the momentum space remain constant in time simply because all moments ($\langle p^n \rangle$) are constants in time. Next, the relativistic and GUP terms \eqref{cr1} - \eqref{cr4}, appearing in \eqref{freegup1}, are found to be
\ber
C_{\text{rel}} &=& -\frac{2\,\hbar^4}{m^4c^2}\lim_{x\rightarrow 0} \left(f_{{\hat{p}_0},4}(x) - f_{{\hat{p}_0},1}(x) f_{{\hat{p}_0},3}(x) \right) \nonumber \\
&& + \frac{\hbar^6}{2m^6c^4} \lim_{x\rightarrow 0} \left( - f_{{\hat{p}_0},6}(x) +  f_{{\hat{p}_0},3}^2 (x) \right)\label{n1} \\
 C_{\text{LGUP}} &=& - \frac{12i \alpha\,\hbar^3}{m^2} \lim_{x\rightarrow 0} (f_{{\hat{p}_0},3}(x) - f_{{\hat{p}_0},1}(x) f_{{\hat{p}_0},2}(x) ) \nonumber \\ && + \frac{18\alpha^2\,\hbar^4}{m^2} \lim_{x\rightarrow 0} \left(f_{{\hat{p}_0},4}(x) -  f_{{\hat{p}_0},2}^2 (x) \right)  \label{n2} \\
C_{\text{QGUP}} &=& \frac{40\alpha^2\,\hbar^4}{m^2} \lim_{x\rightarrow 0} \left(f_{{\hat{p}_0},4}(x) -  f_{{\hat{p}_0},1} (x) f_{{\hat{p}_0},3} (x) \right) \nonumber \\ && +\frac{200\alpha^4\,\hbar^6}{m^2} \lim_{x\rightarrow 0} \left( - f_{{\hat{p}_0},6}(x) +  f_{{\hat{p}_0},3}^2 (x) \right) \label{n3} \\
C_{\text{LGUP}}^{\text{rel}} &=& -\frac{2i\alpha\,\hbar^5}{m^3c^2} \lim_{x\rightarrow 0} (f_{{\hat{p}_0},5}(x) - f_{{\hat{p}_0},1}(x) f_{{\hat{p}_0},4}(x) ) \nonumber \\  && +\frac{25\alpha^2\,\hbar^8}{2m^6c^4} \lim_{x\rightarrow 0} (f_{{\hat{p}_0},8}(x) -  f_{{\hat{p}_0},4}^2(x) )  \label{n4}
\eer
Note also, since the derivatives of the CF are time independent, therefore various parameters in \eqref{cr1} - \eqref{cr4} are also constants in time. With the above mentioned substitutions of \eqref{n1} - \eqref{n4} in  \eqref{freegup1} we finally get the results defined in terms of characteristic functions of momentum operator evaluated in the position space.

\section{GUP based expansion rates with Gaussian wave-packets} \label{sec5}

In this section, we first apply our formalism
to the well-studied Gaussian distribution, and 
calculate the expansion rate. 
Our aim is to verify two things 
--- (i) that the term $\dot{\xi}(t)$ appearing in various formulas vanishes for this case, and (ii) that we now have an alternative way of calculating the various moments appearing in our results by considering - (a) standard quantum mechanical expectation values, or (b) by using the characteristic function. Both lead us to the same conclusion. 
This is quite a powerful result since it implies that one is able to work simply with the distribution that the many particle quantum mechanical system may exhibit. It completely bypasses the standard quantum mechanical calculations of expectation values of various operators, and instead uses directly the statistical mechanical tools based on characteristic functions functions to give results of the wave-packet broadening rates. 
To our knowledge, this strategy has not been used 
before for estimations using 
wave-packets, with or without GUP.

\subsection{Expansion rate using quantum mechanical expectation values}

Let us consider a normalized Gaussian wave-packet of the form
\begin{equation}
    \psi(x) = \frac{1}{(2\pi\xi)^{1/4}}\exp\left(\frac{i}{\hbar}p_0x - \frac{(x-x_0)^2}{4\xi}\right),
    \label{gpsi}
\end{equation}
which represents a minimum wave-packet with $\langle x \rangle = x_0$, $\langle p \rangle = p_0$, ${\Delta x}^2 = {\langle x^2 \rangle - \langle x \rangle^2} = \xi$, and $\Delta p = \frac{\hbar}{2\sqrt{\xi}}$. Its Fourier transformation in momentum space is
\begin{eqnarray}
 \phi(p) &=& \frac{1}{\sqrt{2\pi\hbar}}\int_{-\infty}^{+\infty}dx\, \psi(x) e^{-ipx/\hbar} \\
 &=& \left(\frac{2\xi}{\pi\hbar^2}\right)^{1/4}\exp\left({-\frac{ix_0}{\hbar}(p-p_0) - \frac{(p-p_0)^2 \xi}{\hbar^2}}\right)\label{php}
\end{eqnarray}

In our results, we have moments of $p$ upto the fourth order, and using the standard quantum mechanical definition $\langle p^n \rangle =\int_{-\infty}^{+\infty} dp\, \phi^*(p)p^n\phi(p)$ we can calculate various coefficients in \eqref{n1} - \eqref{n4}. The final results for various coefficients are found to be \cite{Das:2021yqn},
\begin{eqnarray}
 C_{\text{rel}} &=& \frac{3 \hbar ^2}{128 c^4 m^6 \xi_{\text{in}}^3} \left(-16 c^2 m^2 \xi_{\text{in}} \left(4 \overline{p}_0^2 \xi_{\text{in}} + \hbar^2 \right) \right. \nonumber \\
 && \left. + 48 \overline{p}_0^2 \xi_{\text{in}} \hbar^2 + 48 \overline{p}_0^4 \xi_{\text{in}}^2 + 5 \hbar^4\right) \label{crgau}\\
 C_{\text{LGUP}} &=& \frac{3 \alpha  \hbar ^2 }{4 m^2 \xi_{\text{in}}^2} \left(3 \alpha  \hbar ^2+24 \alpha  \overline{p}_0^2 \xi_{\text{in}}-8 \overline{p}_0 \xi_{\text{in}}\right) \label{clgau}\\
 C_{\text{QGUP}} &=& \frac{15 \alpha ^2 \hbar ^2}{8 m^2 \xi_{\text{in}}^3} \left(25 \alpha^2 \hbar^4+16 \overline{p}_0^2 \left(15 \alpha ^2 \xi_{\text{in}} \hbar ^2+ \xi_{\text{in}}^2\right) \right. \nonumber \\
 && \left. +240 \alpha ^2 \overline{p}_0^4 \xi_{\text{in}}^2+4 \xi_{\text{in}} \hbar^2\right) \label{cqgau}\\
 C_{\text{LGUP}}^{\text{rel}} &=& \frac{\alpha  \hbar^2}{16 c^4 m^6 \xi_{\text{in}}^4} \left[8 c^2 m^3 \overline{p}_0 \xi_{\text{in}}^2 \left(4 \overline{p}_0^2 \xi_{\text{in}} +3 \hbar^2\right) \right. \nonumber\\
 && \left. +25 \alpha  \left(84 \overline{p}_0^4 \xi_{\text{in}}^2 \hbar^2+48 \overline{p}_0^2 \xi_{\text{in}} \hbar ^4 \right.\right. \nonumber\\
 && \left.\left. +32 \overline{p}_0^6 \xi_{\text{in}}^3+3 \hbar ^6\right)\right]. \label{clrgau} \nonumber
\end{eqnarray}
Notice that, since $\langle p^n \rangle$ is time independent and all $C_i$'s are expressed in terms of these moments, we can choose any given time to fix these constants. In our case, 
we have chosen the initial value $t_0$ when the packet was minimum 
(i.e. 
$\xi_{\text{in}} = \xi (t=t_0)$).

\subsection{Expansion rate using the characteristic function}
Now, let us calculate various coefficients appearing in the expansion rate using the characteristic function and their definitions in \eqref{n1}, \eqref{n2}, \eqref{n3} and \eqref{n4}. The characteristic function corresponding to the Gaussian wave-packet \eqref{gpsi} is given by \cite{Reichl}
\begin{equation}
    f(x) = e^{i\frac{p_0x}{\hbar} - \frac{x^2}{8\xi_0}}.
    \label{chg}
\end{equation}
It can be shown that $f(x) = \int_{-\infty}^{+\infty} e^{ipx}P(p)$, where, $P(p) = \phi^*(p)\phi(p)$ where $\phi(p)$ is given by \eqref{php}. It is now straightforward to take various derivatives of $f(x)$ and then use the definitions \eqref{n1}-\eqref{n4} to verify that they reproduce \eqref{crgau}, \eqref{clgau}, \eqref{cqgau} and \eqref{clrgau}. 
This completes our 
verification that one can either use the standard definition of quantum mechanical expectation values  or the CFs, contemplating with identical results on the expansion rate for the free wave-packets.

\section{Wave-packets beyond the Gaussian}
\label{sec6}
The analysis in the last section revealed a useful point 
- one can replace the quantum mechanical expectation values for calculating various 
averages simply by computing derivatives of characteristic functions. 
The resulting expressions for the various coefficients turned out to be exactly the same, as they should. 
We will use this insight to generalise the wave-packet broadening results for arbitrary distributions. In principle, one can chose an arbitrary distribution with \emph{finite variance} and implement our prescriptions. In this section, 
we shall consider two such examples -- (i) the Poisson distribution, and (ii) the generalized normal distribution.

\subsection{Poisson Distribution}

We shall start by considering the Poisson distribution. There are a number of reasons why such discrete distributions are relevant to our quantum mechanical set up. 
First of all, one can measure positions and all other observable with finite accuracies. 
%
%
Also, as predicted by GUP, there is an fundamental discreteness of position at the Planck scale or larger
scales.
%
This can be made more precise as follows. We know that while the wavefunctions $\psi(x)$ and $\phi(p)$ are normally regarded as continuous function of its arguments, with  $|\psi|^2$ and $|\phi|^2$ the probability densities in position and momentum space respectively, there are fundamental limitations to distinguishing $\psi(x)$ and $\psi(x+\Delta x)$ or $\phi(p)$ and $\phi(p+\Delta p)$ when $\Delta x$ or $\Delta p$ are below the threshold of measurement accuracy. 
%
%
Although accuracies are continually improving, the best distance intervals that can be measured currently are of the order of $10^{-10}$ m in an electron microscope and $10^{-20}$ m in the Large Hadron Collider. 
Similar bounds exist for the momentum.
Note also that theories with GUP predict a minimum measurable length which is fundamental in nature, independent of measurement accuracies, and the accuracy of 
measurement is limited by it as well. 
This is related to the fact that most theories of 
quantum gravity have a fundamental unit of length, namely the Planck length, about $10^{-35}$ m, which is the minimum measurable length possible, and which cannot be surpassed by any measuring apparatus, however precise. Therefore, from the experimental point of view, a discrete probability distribution is strictly speaking necessary, 
and the Poisson distribution is perhaps the simplest one among them. 
It has just one discrete parameter and only one non-zero 
moment which determines all its properties. 
Therefore we study it as a relevant and interesting example.

We remind that similar to the Gaussian distribution, 
Poisson is also infinitely divisible and has a finite variance. Its Characteristic function is given by
\cite{Reichl}
\begin{eqnarray}
  f_{\hat{p}}(x) = \exp{[i\,\frac{p_0 x}{\hbar} + \frac{\sigma^2}{\delta_p^2}
  (e^{i \frac{x \delta_p}{\hbar}} - 
  \frac{i x \delta_p}{\hbar} - 1)]}
  \label{cfp}
\end{eqnarray}
which corresponds to the Poisson probability density function for the stochastic variable $\hat{P}$, given by
\begin{equation}
    P_{\hat{p}} (p) = \sum_{m=0}^{\infty} \frac{\sigma^{2m}e^{-\sigma^2/\delta_p^2}}{m!\, \delta_p^{2m} } \delta(p-p_0-m\delta_p),
\end{equation}
where $\langle p \rangle = p_0 + \frac{\sigma^2}{\delta_p^2}$ is the average value of the momentum operator and $\sigma$ is the standard deviation of momentum distribution and $\delta_p$ is the spacing between realizations of its stochastic outcome. Note that the relationship between the standard deviation in the momentum space is related to the standard deviation in position space as $\sigma^2=\frac{\hbar^2}{4\xi_0}$.

Using \eqref{cfp} in the expressions \eqref{n1}, \eqref{n2} \eqref{n3} and \eqref{n4}, we can calculate the various coefficients arising in the formula for the expansion rates of the free wave-packet \eqref{freegup1}. After some 
algebra, we finally arrive at the following results
\begin{widetext}
\begin{eqnarray}
C_{\text{rel}} &=& \frac{1}{128 c^4 m^6 \xi_{\text{in}}^3} \left( -16 \xi_{\text{in}}^2 \hbar^2 \left(\delta ^2+3 {p_0}^2+3 \delta  {p_0}\right) \right. \nonumber \\
&&\left. \left(4 c^2 m^2-\delta ^2-3 {p_0}^2-3 \delta {p_0}\right) + 24 \xi_{\text{in}} \hbar^4  \right.\nonumber \\ 
&&\left. \left(-2 c^2 m^2 +4 \delta ^2+6 {p_0}^2+9 \delta  {p_0}\right)+15 \hbar ^6\right) \label{c0kp}\\
C_{\text{LGUP}} &=& \frac{9 \alpha ^2 \hbar ^4+6 \alpha  \xi_{\text{in}} \hbar ^2 (\delta +2 {p_0}) (3 \alpha  \delta +6 \alpha  {p_0}-2)}{4 m^2 \xi_{\text{in}}^2}\\
C_{\text{QGUP}} &=& \frac{15 \alpha ^2 \hbar ^2 \left(25 \alpha ^2 \hbar ^4+4 \xi_{\text{in}} \hbar ^2 \left(60 \alpha ^2 {p_0}^2+1\right)+16 {p_0}^2 \xi_{\text{in}}^2 \left(15 \alpha ^2 {p_0}^2+1\right)\right)}{8 m^2 \xi_{\text{in}}^3} \label{c1kp}\\
C_{\text{LGUP}}^{\text{rel}} &=& \frac{\alpha  \hbar ^2}{32 c^4 m^6 \xi_{\text{in}}^4} \left[8 c^2 m^3 \xi_{\text{in}}^2 \left(2 y^2 (\delta +2 {p_0}) \left(\delta ^2+2 {p_0}^2+2 \delta  {p_0}\right)+\hbar ^2 (5 \delta +6 {p_0})\right) \right. \nonumber \\
&&\left. +25 \alpha  \left(2 \xi_{\text{in}}^2 \hbar ^2 \left(59 \delta ^4  +336 \delta ^2 {p_0}^2+256 \delta  {p_0}^3+84 {p_0}^4+220 \delta ^3 {p_0}\right) \right. \right. \nonumber\\
&&\left.\left. +\xi_{\text{in}} \hbar ^4 \left(121 \delta ^2+96 {p_0}^2+204 \delta  {p_0}\right) +4 \xi_{\text{in}}^3 (\delta +2 {p_0})^2 \left(\delta ^2+2 {p_0}^2+2 \delta  {p_0}\right)^2+6 \hbar ^6\right)\right]. \label{c2kp}
\end{eqnarray}
\end{widetext}
The above expressions provide us a generalization of our earlier results for the Gaussian distribution to the Poisson distribution. 
Note that as expected, in the limit $\delta_p\rightarrow 0$,  the above coefficients reduce the ones for a Gaussian distribution, namely equations  \eqref{crgau}-\eqref{clrgau}.

\subsection{Generalized normal distribution}
Probability density function for the generalized normal distribution in momentum space is given by \cite{tibor}
\begin{equation}
    P_{\hat{p}} (p) = \frac{\kappa}{2\sigma_p \Gamma[1/\kappa]} \exp[-\text{}
\left|    \frac{p-p_0}{\sigma_p} \right|^\kappa]
\end{equation}
where $\kappa,\sigma_p > 0$, $p_0 \in \mathrm{R}$ and $p \in \mathrm{R}$. The standard deviation $\sigma_p$ is also in momentum space. The corresponding CF for this distribution can be calculated following \cite{tibor}, which in this case has the following form
\begin{equation}
    f_{GN} (x) = \frac{\sqrt{\pi}e^{ip_0 x/\hbar}}{\Gamma[1/\kappa]} \sum_{n=0}^{\infty}\frac{\Gamma[1/\kappa + 2n/\kappa]}{\Gamma[1/2+n]} \frac{[-(\sigma_p x/2\hbar)^2]^n}{n!}. \label{cfggn}
\end{equation}
%
One can check that for the specific value of $\kappa=2$, the above reduces to the characteristic function for the normal distribution,
\begin{eqnarray}
    f_{GN} (x,\kappa=2) &=& e^{ip_0 x/\hbar} \sum_{n=0}^{\infty} \frac{[-(\sigma_p x/2\hbar)^2]^n}{n!}\\
    &&= e^{ip_0 x/\hbar} e^{-(\sigma_p x/2\hbar)^2}
\end{eqnarray}
which is precisely \eqref{chg}.

We are now in a position to calculate the various coefficients, using the CF \eqref{cfggn},  appearing in the key expression \eqref{freegup1}. After a bit of algebra we arrive at the following expressions
\begin{eqnarray}
C_{\text{rel}} &=& \frac{\hbar ^2 \Gamma \left({1}/{\kappa }\right)}{16 c^4 m^6 \xi ^3 \Gamma \left({1}/{\kappa }\right)^2} \left[12 \xi ^2 p^2 \Gamma \left({3}/{\kappa }\right) \left(3 p^2-4 c^2 m^2\right) \right.\nonumber\\
&& \left. +2 \xi  \hbar ^2 \Gamma \left({5}/{\kappa }\right) \left(15 p^2-4 c^2 m^2\right)+\hbar ^4 \Gamma \left({7}/{\kappa }\right)\right]  - \frac{9 p^2 \hbar ^4 \Gamma \left({3}/{\kappa }\right)^2}{8 c^4 m^6 \xi^2 \Gamma \left({1}/{\kappa }\right)^2}, \label{c0ggn}\\
C_{\text{LGUP}} &=& \frac{3 \alpha  \hbar ^2 \left(\Gamma \left({1}/{\kappa }\right) \left(3 \alpha  \hbar ^2 \Gamma \left({5}/{\kappa }\right)+8 \xi  p \Gamma \left({3}/{\kappa }\right) (3 \alpha  p-1)\right)-3 \alpha  \hbar ^2 \Gamma \left({3}/{\kappa }\right)^2\right)}{2 m^2 \xi ^2 \Gamma \left({1}/{\kappa }\right)^2}, \label{clggn}\\
C_{\text{QGUP}} &=& \frac{5 \hbar ^2}{m^2 \xi ^3 \Gamma \left({1}/{\kappa }\right)^2} \left(\alpha ^2 \Gamma \left({1}/{\kappa }\right) \left(5 \alpha ^2 \hbar ^4 \Gamma \left({7}/{\kappa }\right)+12 \xi ^2 p^2 \Gamma \left({3}/{\kappa }\right) \left(15 \alpha ^2 p^2+1\right) \right.\right. \nonumber\\
&& \left.\left. +2 \xi  \hbar ^2 \Gamma \left({5}/{\kappa }\right) \left(75 \alpha ^2 p^2+1\right)\right) -90 \alpha ^4 \xi  p^2 \hbar ^2 \Gamma \left({3}/{\kappa }\right)^2\right), \label{cqggn}\\
C_{\text{LGUP}}^{\text{rel}} &=& \frac{\alpha  \hbar ^2}{32 c^4 m^6 \xi ^4 \Gamma \left(\frac{1}{\kappa }\right)^2} \left(\Gamma \left({1}/{\kappa }\right) \left(16 \xi ^2 p \hbar ^2 \Gamma \left({5}/{\kappa }\right) \left(4 c^2 m^3+425 \alpha  p^3\right) \right. \right. \nonumber \\
&& \left. \left. +128 \xi ^3 p^3 \Gamma \left({3}/{\kappa }\right) \left(c^2 m^3+25 \alpha  p^3\right)+25 \alpha  \hbar ^4 \left(\hbar ^2 \Gamma \left({9}/{\kappa }\right)+56 \xi  p^2 \Gamma \left({7}/{\kappa }\right)\right)\right) \right. \nonumber\\ 
&& \left. -25 \alpha  \left(\hbar ^3 \Gamma \left({5}/{\kappa }\right)+12 \xi  p^2 \hbar  \Gamma \left({3}/{\kappa }\right)\right)^2\right). \label{c2ggn}
\end{eqnarray}

It is again possible to cross check the correctness of above results by considering the limit of the Gaussian distribution by setting $\kappa=2$ in the above set of equations \eqref{c0ggn}-\eqref{c2ggn}, which then lead to the corresponding expressions \eqref{crgau}-\eqref{clrgau}. 
This further emphasises the validity and strength of this method in generalising the results of wave-packet expansion for an arbitrary distribution (with finite variance). We now move on to the next section where we present a numerical analysis and study the likelihood of realising these GUP effects in the laboratory for various distributions.

\section{Doubling time difference for various distributions: a numerical study}
\label{sec7}

In this section, we numerically compute the  wave-packet ``doubling times'' for various physically realizable distributions described in the previous section, taking into account the Planck scale corrections. 
We shall study a particular set of molecular wave-packets known as ``Buckyballs''-- $C_{60}$ and $C_{176}$, as well wavepckets of Large Organic Molecular (LOM). These wavepackets were also considered in our earlier 
strictly nonrelativistic
studies \cite{Villalpando:2018xsh, Villalpando:2019usm}, as well as one with the 
leading order relativistic corrections
\cite{Das:2021yqn}, since they are known to  
exhibit quantum mechanical behavior, 
as seen for example, in double slit experiments \cite{dbs1,dbs2,dbs3}. 
It was shown that a precision measurement of time in which a wave-packet of a given initial size doubles its size may be useful in distinguishing between the HUP and GUP based results. However, the above analyses \cite{Villalpando:2018xsh, Villalpando:2019usm, Das:2021yqn} were carried out for the Gaussian distribution. Given the general 
formalism we have developed for the potential measurement of Planck scale effects, we now test its robustness 
with more general distributions. 
While our prescription works for \emph{any} distribution with finite variance, we shall consider two concrete cases for our numerical study, namely wavepackets following (i) Poisson and (ii) generalized normal distributions.  

We define the doubling time difference (DTD) in the following manner,
\begin{equation}
  \Delta t_{\text{double}} =  t^{\text{GUP}}_{\text{double}} - t^{\text{HUP}}_{\text{double}},
  \label{DTD}
\end{equation}
where the two terms on the right hand side signify the times required for a free wavepacket to double its width following the GUP and HUP expansion rates respectively. 
While $t^{\text{GUP}}_{\text{double}}$ in the above can be eaily obtained from the general expression \eqref{freegup1} which incorporates both GUP and relativistic effects and their mutual interplay, $t^{\text{HUP}}_{\text{double}}$ can obtained from 
the same expression by setting 
the GUP coefficients ($C_{\text{LGUP}}$, $C_{\text{QGUP}}$ and $C_{\text{LGUP}}^{\text{rel}}$) to zero, while still retaining the pure relativistic terms.
%
Then equation \eqref{DTD} provides us an estimate of the DTD, for a given distribution, in presence of the Planck scale modifications. In the following  we consider the aforementioned readily available molecular wavepackets and see how distributional variations can affect the DTD for each of those case studies.

\subsection{Buckyball C-60 molecular wave-packet}
It is well known that the C-60 ``Buckyball'' molecule exhibits interference patterns in a double slit experiment \cite{dbs1}. Considering this wavepacket as a whole, we obtain the following values of the relevant physical parameters, namely mass $m=1.1967 \times 10^{-24}$ kg and initial width $\Delta_{\text{in}} = 7 \times 10^{-10}$ m. In addition we shall assume the mean velocity of the wavepacket $v=10^5$ m/s, which is a characteristic value that renders both the GUP and relativistic terms non-negligible and makes our formalism applicable. 

With the above in place, we can now play with various distributions, by considering GUP coefficients for the Gaussian \eqref{crgau}-\eqref{clrgau} or the  Poisson \eqref{c0kp}-\eqref{c2kp} or the generalized normal distributions \eqref{c0ggn}-\eqref{c2ggn}. Recall also that none of those coefficients survive when we consider the HUP limit, and therefore it is straightforward to numerically compute the DTD as defined in \eqref{DTD}. Note also that for the generalized normal distribution \eqref{c0ggn}-\eqref{c2ggn} we have a parameter $\kappa$, and for $\kappa=2$ we get back the standard Gaussian limit which corresponds to our expressions \eqref{crgau}-\eqref{clrgau}.

\begin{figure*}[h]
    \centering
    \begin{subfigure}
        \centering
        \includegraphics[width=0.45\linewidth]{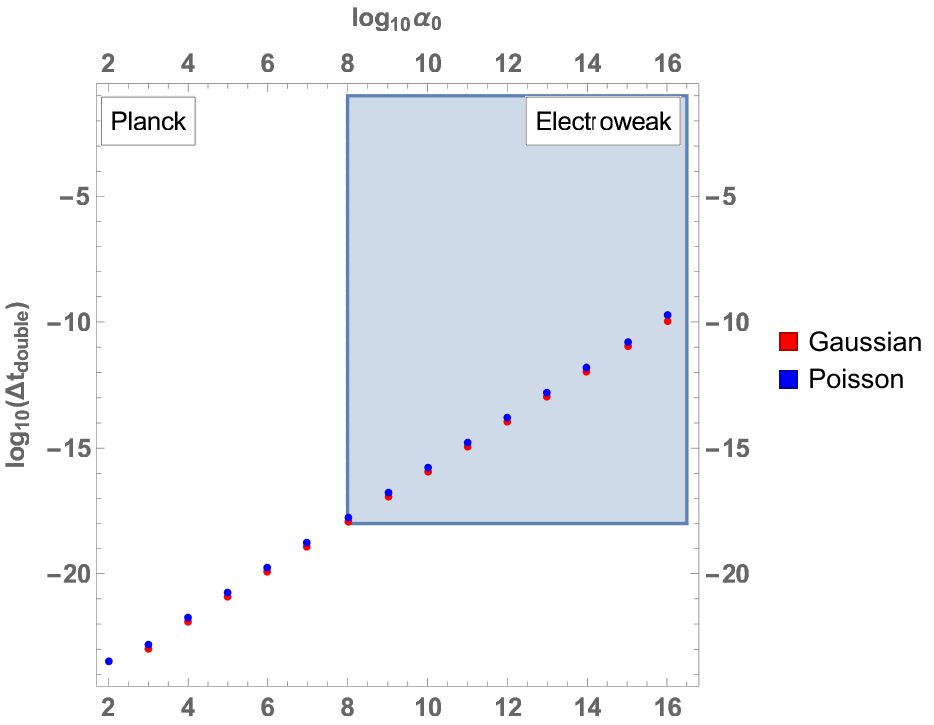}
    \end{subfigure}
    \begin{subfigure}
        \centering
        \includegraphics[width=0.45\linewidth]{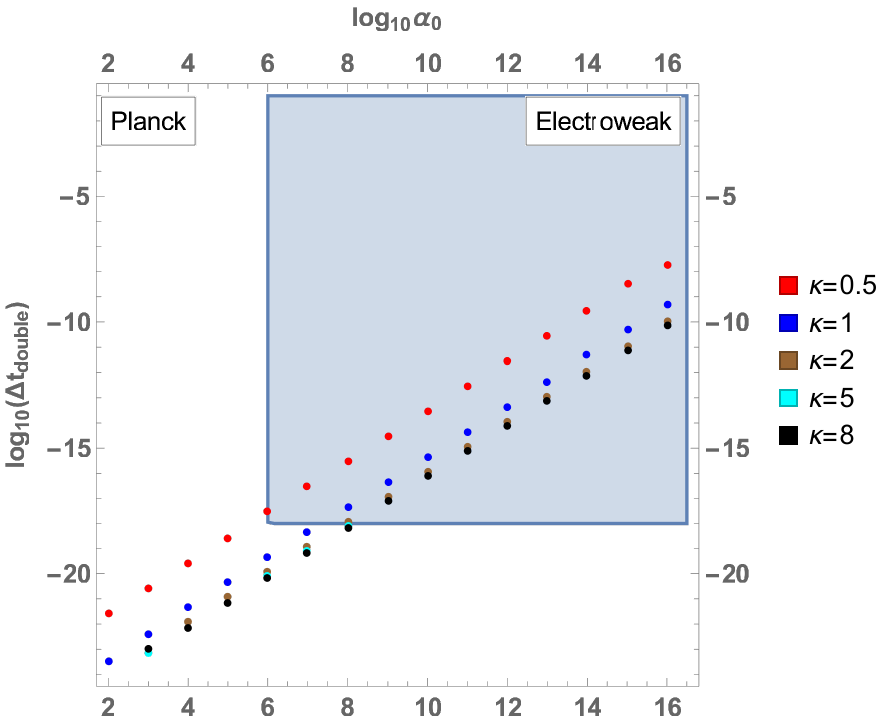}
    \end{subfigure}
    \caption{\label{fig-1} Doubling time difference versus the GUP parameter plot in a logarithmic scale using our numerical analysis fot C-60 ``buckyball'' molecule. The left panel of the figure corresponds to the results for the Gaussian versus Poisson distributions. We can see that an equivalent parameter space can be probed by both distributions. The right panel is the gist of results obtained by considering the generalized normal distribution with five different values of the parameter $\kappa$. While $\kappa=2$ gives us the Gaussian distribution, for all other values it differs with Gaussian. We see that lower values of the parameter $\kappa<2$ enhance our chances of measuring the doubling time difference. For $\kappa=0.5$, and  with an atomic clock of attosecond ($10^{-18}$s) accuracy, one is able to scan the parameter space which is equivalent to 
    $10^6\times l_{\text{Planck}}$.}
\end{figure*}

The summary of our numerical analysis is plotted using a logarithmic scale in Figure-\ref{fig-1}. The left panel of the figure shows a comparison of the DTD with the GUP parameter ($\alpha$) for the Gaussian and Poisson distributions, while the right panel summarizes results for generalized normal distributions for several choices of the parameter $\kappa$. These plots give us useful information on how the Planck scale effects would be manifested according to the distribution of the wavepacket. For example, 
changing the Gaussian to Poisson slightly increases the likelihood of detecting DTD.
However, there is a significant variation in DTD for generalized normal distribution that depends on the parameter $\kappa$. It appears to be a general trend that lower the value of $\kappa$, higher the DTD corresponding to a given value of $\alpha$. We can make a comparison keeping the standard Gaussian distribution ($\kappa=2$) as a reference, which then shows us the DTD indeed gets enhanced for values of $\kappa<2$. We are able to calculate numerically for the case $\kappa=0.5$ which gives considerably  better DTD than the Gaussian distribution. We are able to test the GUP parameter space ($\alpha$) down to one order of magnitude lower than the Gaussian case. Therefore, we conclude that selecting an appropriate distribution may be crucial in testing these quantum gravity corrections by means of time resolved experiments.  

Next, we show that these results can me made even better by choosing bigger molecular wavepackets. We shall consider two such cases in the following subsections.

\subsection{Buckyball C-176 molecular wave-packet}

It is also well known that the C-176 ``Buckyball'' molecule exhibits interference patterns in a double slit experiment \cite{dbs2}. Considering this wavepacket as a whole, we obtain the following values of the relevant physical parameters -- mass $m=3.5070 \times 10^{-24}$ kg and initial width $\Delta_{\text{in}} = 1.2 \times 10^{-9}$ m. Like before, we shall assume the mean velocity of the wavepacket $v=10^5$ m/s, which again gives rise to both Planck scale as well as leading order relativistic corrections. 

The summary of results is plotted in Figure \ref{fig-2}. Once again, we find that keeping everything but the distribution unaltered leads one to the previous conclusion -- that the generalized distribution with $\kappa<2$ is indeed preferable over the Gaussian distribution for testing these Planck scale modifications. 

\begin{figure*}[h]
    \centering
    \begin{subfigure}
        \centering
        \includegraphics[width=0.45\linewidth]{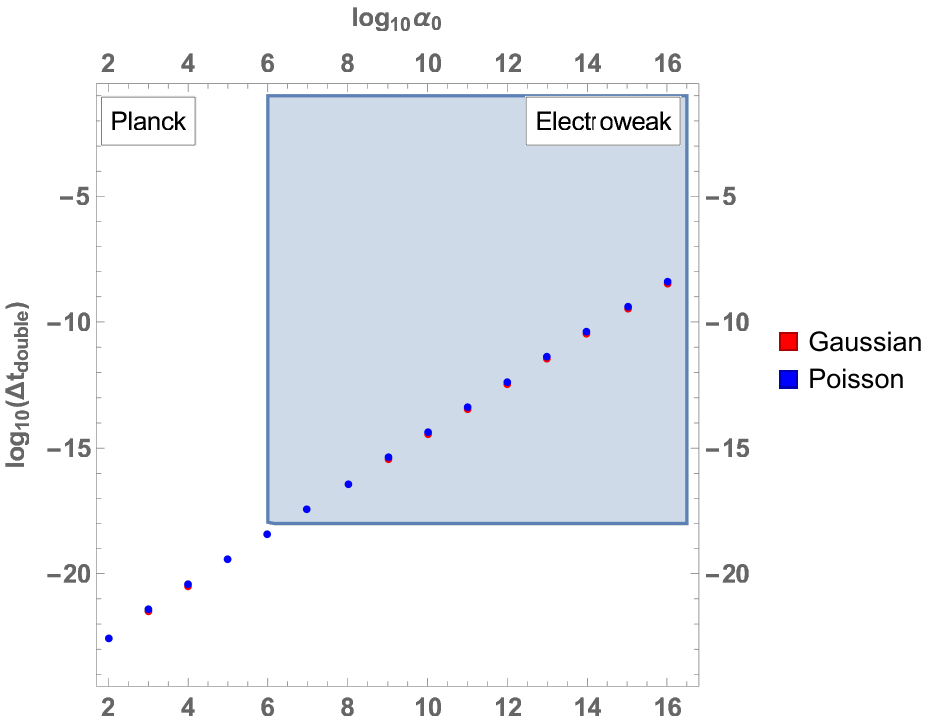}
    \end{subfigure}
    \begin{subfigure}
        \centering
        \includegraphics[width=0.45\linewidth]{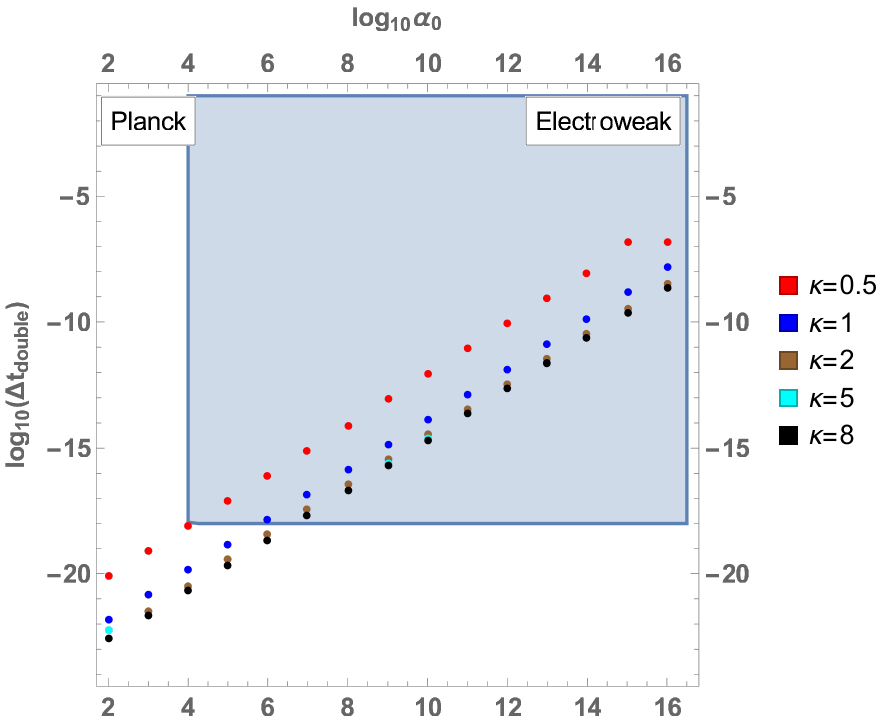}
    \end{subfigure}
    \caption{\label{fig-2} Doubling time difference versus the GUP parameter plot in a logarithmic scale using our numerical analysis for C-170 ``buckyball'' molecule. The left panel of the figure corresponds to the results for the Gaussian versus Poisson distributions. We see that an equivalent parameter space can be probed by both distributions. The right panel is the summary of results obtained by considering the generalized normal distribution with five different values of the parameter $\kappa$. While $\kappa=2$ gives us the Gaussian distribution, for all other values it differs from the Gaussian. We see that lower values of the parameter, i.e. $\kappa<2$ increases the chance of measuring the doubling time difference. For $\kappa=0.5$, and  with an atomic clock of attosecond ($10^{-18}$s) accuracy, one is able to scan the parameter space which is equivalent to 
    $10^4\times l_{\text{Planck}}$.}
\end{figure*}

\begin{figure*}[t]
    \centering
    \begin{subfigure}
        \centering
        \includegraphics[width=0.45\linewidth]{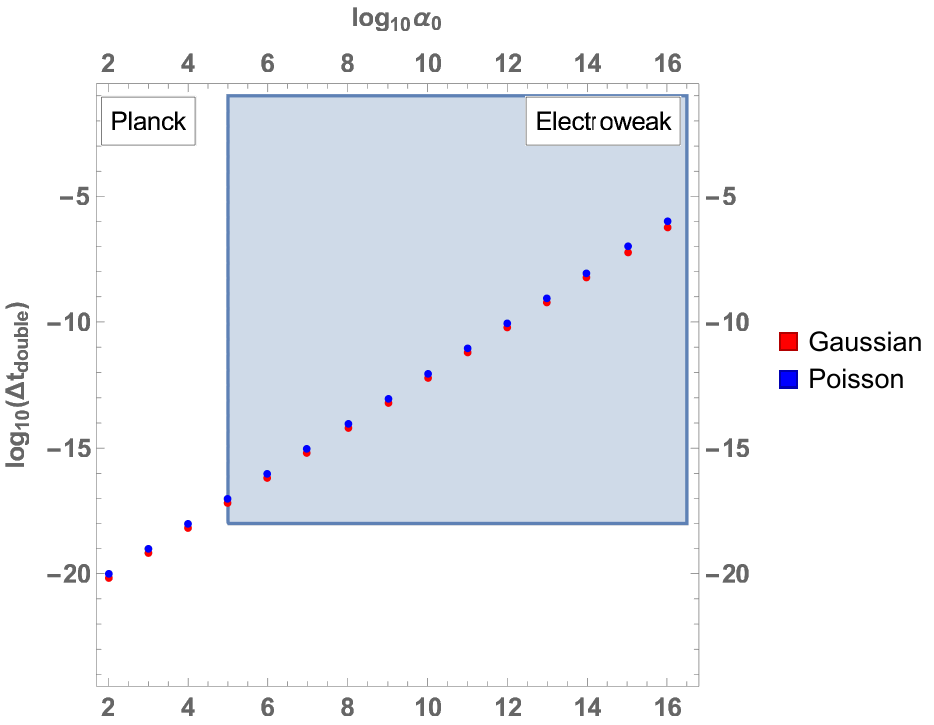}
    \end{subfigure}
    \begin{subfigure}
        \centering
        \includegraphics[width=0.45\linewidth]{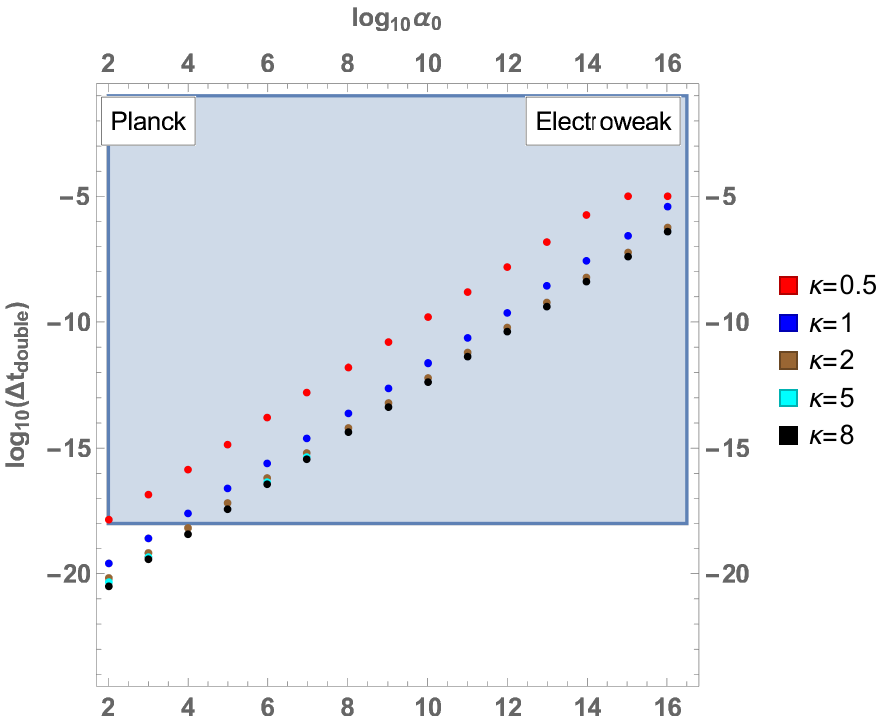}
    \end{subfigure}
    \caption{\label{fig-3} Doubling time difference versus the GUP parameter plot in a logarithmic scale using our numerical analysis for the TPPF152 molecule. The left panel of the figure corresponds to the results for the Gaussian versus Poisson distributions. We can see that an equivalent parameter space can be probed by both distributions. The right panel is the gist of results obtained by considering the generalized normal distribution with five different values of the parameter $\kappa$. While $\kappa=2$ gives us the Gaussian distribution, for all other values it differs with Gaussian. We see that lower values of the parameter $\kappa<2$ increases the chance of measuring the doubling time difference. For $\kappa=0.5$, and  with an atomic clock of attosecond ($10^{-18}$s) accuracy, one is able to scan the parameter space which is equivalent to 
    $10^2\times l_{\text{Planck}}$. With a clock slightly better, such as zeptosecond ($10^{-20}$s) accuracy, one may be able to go down right to the Planck length. However, our numerical precision compels us to stay slightly above the Planck value.}
\end{figure*}

\subsection{Large Organic Molecular (LOM) wave-packet}
Large-Organic-Molecular TPPF152 wavepacket was constructed a decade ago and it clearly showed quantum behavior due to the expected interference patterns \cite{dbs3}. Considering this wavepacket as a whole, we obtain the following values of the relevant physical parameters -- mass $m= 8.8174\times10^{-24}$ kg and initial width $\Delta_{\text{in}} = 6\times 10^{-9}$ m. Like the previous two cases, we shall assume the mean velocity of the wavepacket $v=10^5$ m/s, which is a characteristic value following the inclusion of leading order relativistic correction. Summary of results corresponding to the numerical study is plotted in Figure \ref{fig-3} which again gives an identical physical scenario like two previous cases. It is also evident that of all the variations, in terms of the choice of molecules and distributions, the most preferable in terms of showing Planck scale modification is the TPPF152 molecular wavepacket with a generalized Gaussian distribution with $\kappa=0.5$. Indeed, such a clear and strong perception on the choice of quantum systems might take us one step further while designing a potential test of such tiny quantum gravity effects in a laboratory. Of course, one might come up a better combination for the choice of molecular wavepacket and its  distribution, and most importantly, our general construction in this work would allow to test all such combinations and might be an useful tool for future experiments.

\section{Conclusions}\label{sec8}

In this paper, we have made key improvements on our earlier works \cite{Villalpando:2018xsh}-\cite{Das:2021yqn} estimating quantum gravity/GUP effects by studying the ``doubling time'' of free wavepackets.  The difference between the doubling times (i.e., based on the standard quantum mechanical and GUP based approaches) were shown to be dependent on the mass, initial size and mean velocity of the wavepacket. However, all the calculations were performed for a Gaussian wavepacket. In this paper, we built an elaborate framework where QG/GUP effect can be computed irrespective of any shape of the wavepacket. We numerically calculated the effect of various shapes on the doubling time differences and concluded that they are important to be considered. This is an important result since now we have complete information on the requirements, applicable to the wavepackets, which might be helpful in setting up an appropriate experimental set up for chasing these tiny QG/GUP deviations. 

In order to achieve the above mentioned result we exploited a novel duality between the quantum mechanical and statistical mechanical tools, which finally allows us to consider any shape of the wavepacket following an {\it arbitrary} statistical distribution. The only physical restriction is that the distributions must have a finite variance.  We have shown that the potentially observable quantities, although quantum mechanical in nature, can be computed by purely using machinery of statistical mechanics, namely that of characteristic functions. To test this, we considered a couple of distributions beyond the Gaussian, namely the Poisson and the generalized normal distributions. We computed the corresponding characteristic functions, wrote the doubling times in terms of them, and finally estimated the times numerically, for a variety of laboratory based systems, namely the buckyball and the large organic molecules. Remarkably, we found that the doubling times are still tiny, but just within current measurement accuracies. This shows the importance of choosing the right distribution with the right parameters in being able to detect the otherwise tiny minimal scale effects. The actual detection of the above effect would point towards new physics at microscopic length scales, and will likely
shed important light on quantum gravity theories and the fundamental nature of spacetime itself. 
On the other hand, even the absence of such detection would significantly improve the ever-tightening bounds on the minimal scale parameters. Either way, this shows the importance of looking for such effects in laboratory based systems such as described here. Our numerical results with LOM wavepackets with generalized normal distribution $\kappa=0.5$ shows measurable time difference which corresponds to a minimal length scale hundred times the Planck value. Therefore, our strategy has capability to scan the minimal length scale post GUT scale and almost near the Planck scale.



\section*{Acknowledgements}
This work is supported by the Natural Sciences and Engineering Research Council of Canada 
and the Alberta Government Quantum Major Innovations Fund. 
Research of SKM is supported by SEP-CONACyT research grant CB/2017-18/A1S-33440, Mexico and CONACyT ``Ciencia-Frontera'' grant 140630.

Data Availability Statement: No Data associated in the manuscript.

\end{document}